\documentclass[onefignum,onetabnum]{siamsials251208}

\usepackage{lipsum}
\usepackage{amsfonts}
\usepackage{graphicx}
\usepackage{epstopdf}
\usepackage{algorithmic}
\ifpdf
  \DeclareGraphicsExtensions{.eps,.pdf,.png,.jpg}
\else
  \DeclareGraphicsExtensions{.eps}
\fi

\usepackage{enumitem}
\setlist[enumerate]{leftmargin=.5in}
\setlist[itemize]{leftmargin=.5in}

\newsiamremark{remark}{Remark}
\newsiamremark{hypothesis}{Hypothesis}
\crefname{hypothesis}{Hypothesis}{Hypotheses}
\newsiamthm{claim}{Claim}
\newsiamremark{fact}{Fact}
\crefname{fact}{Fact}{Facts}

\headers{Slow rhythms in delay-coupled oscillators}{X. Qie, M. Martin, S. Liu, and M. G. Pedersen}
\title{Dissecting emerging slow rhythms in delay-coupled neural oscillators \thanks{Submitted to the editors \today. \\ X.~Qie and M.~Martin contributed equally and are joint first authors.
\funding{This work was partly funded by the National Natural Science Foundation of China (Grant Nos. 11872183) and China Scholarship Council (Grant Nos. 202506150077})}}

\author{Xinxin Qie\thanks{School of Mathematics, South China University of Technology, Guangzhou, China
(\email{xinxinqie77@gmail.com}).}
\and Matteo Martin\thanks{Department of Information Engineering, University of Padova, Italy
  (\email{matteo.martin.2@phd.unipd.it}).}
\and Shenquan Liu\thanks{School of Mathematics, South China University of Technology, Guangzhou, China
(corresponding author; \email{mashqliu@scut.edu.cn}).}
\and Morten Gram Pedersen\thanks{Department of Information Engineering, and Padova Neuroscience Center, University of Padova, Italy
(lead corresponding author; \email{mortengram.pedersen@unipd.it}, \url{https://www.dei.unipd.it/~pedersen/}).}
}

\usepackage{amsopn}

\ifpdf
\hypersetup{
  pdftitle={Slow rhythms in delay-coupled oscillators},
  pdfauthor={X. Qie, M. Martin, S. Liu, and M. G. Pedersen}
}
\fi

\usepackage{xspace}

\usepackage[sort&compress]{natbib}
\usepackage{hhline}                 
\usepackage{tablefootnote}          
\usepackage{tabularx}               
\usepackage{booktabs}               
\usepackage{colortbl}               
\usepackage{multirow}               
\usepackage{makecell}               
\setcitestyle{numbers,square,comma}
\usepackage{soul} 

\begin{document}

\maketitle

\begin{abstract}
Synaptic transmission delays are ubiquitous in neural circuits and can alter the dynamical repertoire of coupled oscillators quantitatively and qualitatively. 
Here, we demonstrate that delayed coupling in inhibitory networks introduces an effective slow–fast structure in the phase-difference dynamics, generating low-frequency components that 
are not due to
intrinsic 
cellular properties,
and we show that this behavior is not specific to
a particular model structure.
The origin of this generic phenomenon is analyzed by numerical continuation and bifurcation analysis,
which provides a systematic approach to find such delay-induced slow modulating rhythms.
\end{abstract}
\begin{relevance}
Coupled oscillatory dynamics is a common theme in neural systems, from coordinated cortical rhythms to oscillatory small-scale networks such as central pattern generators. 
Coupling delays, inherent in neural communication due to axonal conduction and synaptic transmission, influence phase relations between the individual neural oscillators and enrich network dynamics. 
The slow modulation investigated in this work suggests that neural circuits can generate patterns with multiple timescales without requiring dedicated slow ionic currents or specialized cellular phenotypes. 
We validate the generality of this phenomenon through three distinct modeling frameworks, ranging from central pattern generators to population-level mean-field models.  
In each case, delayed coupling can induce analogous slow rhythms, demonstrating that this is a robust dynamical consequence of delayed inhibitory coupling rather than a model-specific peculiarity. These results establish that transmission delays provide a generic mechanism for generating slow rhythmic modulation in neural circuits, with implications for motor patterns, phase coordination, and multiscale temporal organization of biological rhythms.
\end{relevance}
\begin{mathcontent}
We employ phase reduction based on phase response curves to derive a phase-difference model with delay for mutually inhibitory coupled oscillators, where the individual units are given by the FitzHugh–Nagumo model, the Morris–Lecar model, or a next-generation neural mass model derived from quadratic integrate-and-fire neurons.
We use phase planes to study multistability and limit cycles, which correspond to slow modulation of fast oscillations in the full model.
Treating the synaptic delay as a bifurcation parameter, we apply numerical continuation to construct delay-dependent bifurcation diagrams. 
The analysis reveals Hopf, heteroclinic, and saddle-node-of-periodics bifurcations that cause and organize slow rhythmic behavior.
Our analysis provides a systematic approach to the search for limit cycles in phase-reduction models corresponding to delay-induced slow rhythms in the original model.
\end{mathcontent}



\section{Introduction}

Rhythmic patterns in neural circuits arise from the interplay of intrinsic
cellular activity and network connectivity \cite{buzsaki2004neuronal}. 
While single neurons can exhibit oscillatory dynamics through the interaction of fast and slow ionic currents, the temporal structure of network-level rhythms may involve timescales that emerge from coupling 
\cite{del2002respiratory,loppini2018gap,phillips2024interdependence}.
Delayed communication between neurons due to synaptic transmission is ubiquitous with delays ranging from milliseconds in local microcircuits to tens of milliseconds across brain regions \cite{lemarechal2022brain}. 
Such delays can fundamentally reshape the dynamical repertoire of coupled oscillators \cite{campbell2012phase,campbell2007time,Niebur1991}.
For example, in central pattern generators (CPGs), the neural circuits underlying, e.g., locomotion, respiration, and swimming \cite{marder2001central,grillner2006biological,ijspeert2008central}, delays provide an essential mechanism for regulating phase relations and enabling the flexible coordination of rhythmic motor patterns \cite{qie2025synaptic,campbell2018phase}.
Similar delay-dependent dynamics operate in thalamic pacemaker synchronization, cortical rhythm generation, and the coordination of neural activity across distributed brain areas \cite{haken2002brain}.

Despite the recognized importance of synaptic delays in neural dynamics \cite{ermentrout1998fine,ermentrout2009delays,laing2009chimera}, it is still not completely clear how delays generate emergent timescales that are absent from the intrinsic dynamics of isolated neurons. 
Previous work, through brute-force model simulations, has established that delays can alter synchronization properties, induce multistability, and destabilize phase-locked states \cite{collens2020dynamics,wojcik2014key,kelley20202}. 
Another interesting consequence of delayed coupling is the emergence of slow modulatory rhythms that have no counterpart in the dynamics of isolated oscillators \cite{qie2025synaptic}. However, the dynamic mechanisms underlying delay-induced slow modulatory rhythms – temporal structures much slower than the intrinsic oscillation period - have not been systematically characterized. 
Moreover, it remains unclear whether such delay-induced slow dynamics represent generic features of coupled neural oscillators, or if they arise only in a few, specific models. 
A theoretical framework that links the delay-coupled model dynamics to slow components in neural behavior would provide important insights into the temporal organization of biological neural circuits. 

To address these questions, we employ phase reduction theory, a mathematical framework that simplifies the analysis of weakly coupled oscillators by describing their interactions through phase variables \cite{ermentrout2010mathematical}. 
In this approach, each neuron is represented by a single phase coordinate, and network interactions are encoded through phase response curves (PRCs) that capture how perturbations shift oscillations \cite{winfree1980geometry,ermentrout1996type,brown2004phase,zhu2026dynamical}.
Crucially for the present work, in the phase model, coupling delays appear explicitly in the phase interaction terms \cite{ermentrout1994introduction,campbell2007time}.
The resulting low-dimensional phase-difference equations retain the essential dynamical features of the full neural delay-coupled models while enabling straight-forward numerical bifurcation analysis. 
We exploit this fact to track how solution branches, representing synchronous states, traveling waves, and non-uniform phase-locked patterns, evolve as the delay varies, by standard numerical continuation. 
Our analysis identifies bifurcations giving rise to limit cycles in the phase model, which correspond to slow modulation of the oscillatory behavior in the full neural model. 
That is, the individual oscillators continue to fire very close to their intrinsic frequency, yet their relative phases evolve along stable limit cycles.
Our approach to finding such slow delay-induced patterns is applicable across diverse neuronal models, since it requires only that the coupling be weak and that the single-unit oscillations be stable.

The paper is organized as follows. In Section 2, we derive the general phase-reduced model for delay-coupled oscillators on a ring network. In Section 3, we apply this framework to an inhibitory circuit using the FitzHugh–Nagumo model, obtaining detailed delay-dependent bifurcation diagrams and identifying slow components in the form of stable limit cycles.
Section 4 validates these findings in the Morris–Lecar model, demonstrating that our approach is applicable to more biophysically realistic models. 
Section 5 extends the analysis to a mean-field neural mass model, establishing that delay-induced slow–fast structures persist at the population level and for models where the single-unit model is of dimension greater than two. 
Section 6 summarizes the main conclusions and discusses the broader implications of our results. 
Overall, we show that transmission delays provide a generic mechanism for generating slow rhythmic modulation in neural systems: by destabilizing equilibria and causing limit cycle dynamics in phase-difference space, delays create effective timescale separation without requiring intrinsic slow variables. This insight bridges low-dimensional dynamical systems analysis with the multiscale temporal structure observed in biological neural circuits, offering new perspectives on how network connectivity shapes rhythmic coordination across the nervous system.

\section{Derivation of the Reduced Phase Model}
We consider a network of $N$ weakly delay-coupled oscillators arranged on a symmetric ring with nearest-neighbor interactions. The state of the $j$-th oscillator, denoted by $\mathbf{X}_j(t) \in \mathbb{R}^n$, evolves according to the delay-differential equation
\begin{equation}\label{dXdt}
\frac{d}{dt} \mathbf{X}_j(t)
=
\mathbf{F}\big(\mathbf{X}_j(t)\big)
+
g \sum_{k=1}^{N} w_{jk}
\, \mathbf{G}\!\left(\mathbf{X}_j(t), \mathbf{X}_k(t-\rho)\right),
\quad j = 1,\dots,N.
\end{equation}
Here, $\mathbf{F}:\mathbb{R}^n \to \mathbb{R}^n$ defines the intrinsic oscillator dynamics, $\rho \geq 0$ denotes the transmission delay, $g>0$ is a small coupling parameter, and $\mathbf{G}:\mathbb{R}^n \times \mathbb{R}^n \to \mathbb{R}^n$ describes the interaction between oscillators. 
The coupling matrix $(w_{jk})$ encodes symmetric nearest-neighbor coupling on a ring, i.e.,
\begin{equation}
w_{jk} =
\begin{cases}
1, & \text{if } j = k \pm 1 \pmod{N},\\
0, & \text{otherwise}.
\end{cases}
\label{EQ_wjk}
\end{equation}
We assume that the isolated oscillators follow a stable limit cycle and apply the theory of weakly coupled oscillators to reduce \eqref{dXdt} to a phase model \cite{ermentrout2010mathematical}.
Let $\theta_j\in \mathbb{S}^1=\mathbb{R}/\mathbb{N}$ be the phase of oscillator $j$ (modulo 1), $T$ the period of the uncoupled oscillator, $\tau=t/T$ a transformed timescale, and let overdot denote differentiation with respect to $\tau$. The phase model, including delay, is then \cite{ermentrout1994introduction}
\begin{equation}
\dot{\theta}_j
=
\omega
+g \sum_{k=1}^{N} w_{jk}
\, H(\theta_k - \theta_j - \eta)
+
\mathcal{O}(g^2),
\quad j=1,\dots,N,
\end{equation}
where $\omega = 1$
and $\eta =  \rho/T$ represents the delay expressed as a phase shift.
The interaction function $H:\mathbb{S}^1 \to \mathbb{R}$ is given by the averaging formula
\begin{equation}\label{eq:H}
H(\vartheta)
=
\int_0^1
\mathbf{Z}(\phi)
\cdot
\mathbf{G}\big(
\mathbf{X}(\phi\, T),
\mathbf{X}((\phi + \vartheta)\, T)
\big)
\, d\phi,
\end{equation}
where $\mathbf{X}(\phi\, T)$ parametrizes the limit cycle of the isolated oscillator, and $\mathbf{Z}(\phi)$ is the associated phase response curve obtained from the adjoint equation.
This approximation reduces the original delay-differential network to a finite-dimensional phase system, with an error of order $\mathcal{O}(g^2)$.

To analyze collective phase-locked states on the ring, we introduce phase differences relative to oscillator~1,
\begin{equation}\label{eq:Delta}
\Delta_{1j} = \theta_1 - \theta_j,
\quad j=2,\dots,N.
\end{equation}
Differentiating and substituting the phase equations, and neglecting $\mathcal O(g^2)$ terms, yield a $(N-1)$-dimensional system governing the relative phase differences,
\begin{equation}\label{eq:dDelta}
\begin{aligned}
\dot{\Delta}_{1j}
=
g\Big[
&H(-\Delta_{12} - \eta)
+ H(-\Delta_{1N} - \eta) \\
&- H(\Delta_{1j} - \Delta_{1,j-1} - \eta)
- H(\Delta_{1j} - \Delta_{1,j+1} - \eta)
\Big],
\end{aligned}
\end{equation}
where $j = 2,3,\dots,N$ and indices are taken modulo $N$.
Critically, $H$ is independent of the delay, which allows us to calculate $H$ as in \eqref{eq:H} only once, and then vary $\eta$ to perform, e.g., bifurcation analysis investigating the effect of delay on the dynamics of the system.

\section{Phase description and delay-induced slow rhythms in the FitzHugh--Nagumo Model}
\label{sec:FHN}
To model the individual neurons, we employ a modified FitzHugh--Nagumo (FHN) model \cite{collens2020dynamics,qie2025synaptic}.  
This model provides a simplified description of the firing dynamics of an excitable neuron and is expressed as a two-dimensional system. We  couple $N$ such FHN neurons in a ring to their nearest neighbors with delayed synaptic transmission,
\begin{equation}\label{dV_FHN}
\begin{split}
\frac{d v_j}{d t}
&=
v_j - v_j^3 - h_j + I + g\sum_{i = 1}^{N}{w_{ji}\,G_{FHN}(v_j(t),v_i(t-\rho))}
,\\
\frac{d h_j}{d t}
&=
\varepsilon\left[\frac{1}{1+e^{-k\,v_j}} - h_j\right],
\end{split}
\end{equation}
for $j = 1,2,\dots,N$, and $w_{ji}$ as in \eqref{EQ_wjk}. The variable $v_j$ denotes the membrane potential, and $h_j$ characterizes the slow recovery variable representing the gating dynamics of ion channels.  
The system exhibits stable limit-cycle oscillations within a suitable range of parameters; we use
$\varepsilon = 0.3$ and $I = 0.4$.  
Here, $I$ corresponds to the external input current, while $\varepsilon$ is a small parameter that ensures a separation of time scales between $v$ and $h$.  
The neurons are connected by unidirectional synapses modeled using the fast threshold modulation (FTM) scheme with delay $\rho$ \cite{qie2025synaptic},
\begin{equation}\label{G_FHN}
G_{FHN}\left(v_j, v_i\right)=
\, \Gamma(v_i)\,\left(v_{\mathrm{rev}}-v_{j}\right),
\end{equation}
where $\Gamma(v)$ is a sigmoidal activation function of the form
\begin{equation}
\Gamma(v_i) = \frac{1}{1+e^{-100\left(v_i-v_{\mathrm{th}}\right)}}.
\end{equation}
The FTM formalism provides a clear distinction between the active (“on”) and inactive (“off”) states of a neuron:  
$\Gamma(v_i) \approx 1$ when the presynaptic voltage $v_i$ exceeds the synaptic threshold $v_{\mathrm{th}}=0$, and $\Gamma(v_i)\approx 0$ otherwise.  
We assume that synapses are inhibitory and therefore set the synaptic reversal potential $v_{\mathrm{rev}}=-1.5$. Finally, $g=0.0005$ is the coupling strength.

\begin{figure}[htbp]
  \centering
  \includegraphics
  {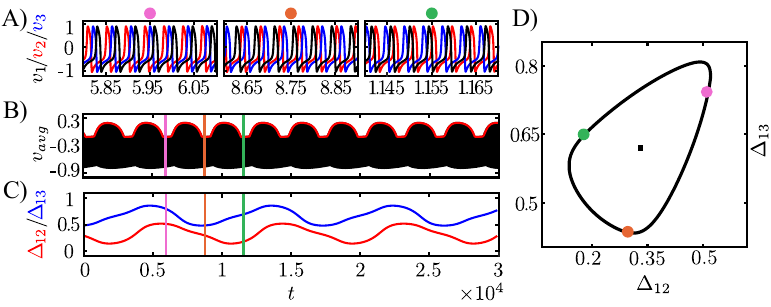} 
  \caption{Delay-induced slow modulation of spiking activity in the FHN model with delay parameter $\eta=0.08$.
  (A) Simulated time series of $v_1$, $v_2$, and $v_3$ (black, red, and blue, respectively) at different time intervals. Note how the relative order of the three units changes between subpanels.
  The pink, orange, and green dots over each subpanel correspond to the vertical bars in panels B and C, and to the circles in panel D.
  (B) Average voltage time course. The red curve shows the the maximum value of the envelope of
  the average voltage.
  (C) Temporal evolution of $\Delta_{12}$ (red) and $\Delta_{13}$ (blue).
  (D) Trajectory of the phase difference model in the $(\Delta_{12}$, $\Delta_{13})$ phase plane. The curves in panels C and D were simulated using the phase difference model, but they are indistinguishable from the phase differences computed from the full model.}
\label{Fig1_FHN3_TS}
\end{figure}

Qie et al.~\cite{qie2025synaptic} showed that three CPG neurons represented by the FHN model with weak delayed coupling, \eqref{dV_FHN} and \eqref{G_FHN}, can exhibit a range of dynamic behaviors, including slow oscillations such as those presented in Figure~\ref{Fig1_FHN3_TS}. 
Specifically, Figure~\ref{Fig1_FHN3_TS}A illustrates the slow modulation of the electrical activity of the single neurons due to weak, delayed coupling, which results in phase shifts among the three neurons, i.e., a recurrent rearrangement of the temporal relationships between the individual voltage traces. 
As a result, a slow rhythm emerges in the average activity (Figure~\ref{Fig1_FHN3_TS}B), which corresponds to slow oscillations in the phase-difference model, as seen in its temporal and phase-plane behaviors (Figure~\ref{Fig1_FHN3_TS}CD).

Based on their numerical observations, the authors \cite{qie2025synaptic} hypothesized the bifurcation scenario as the delay is varied.
We numerically calculated the $H$ function \eqref{eq:H} associated with the limit cycle of the FHN model \eqref{dV_FHN}. 
The $H$ function, which completely defines the 2D phase model \eqref{eq:dDelta} for a given delay $\eta$,  was implemented in XPPAUT \cite{ermentrout02xppaut}. 
This formulation permits standard phase plane analysis, including calculation of nullclines, equilibria and stability analysis. 
Further, it allows numerical continuation using AUTO to directly calculate bifurcation diagrams as the delay $\eta$ varies, which we used to confirm the bifurcation structure proposed by Qie et al.~\cite{qie2025synaptic}. For the visualization of the bifurcation structure of the phase-difference model, we used XPPLORE~\cite{Martin2025}, a flexible toolbox for processing and visualizing XPPAUT continuation results via MATLAB~\cite{MATLAB}.

\begin{figure}[htbp]
    \centering
    \includegraphics{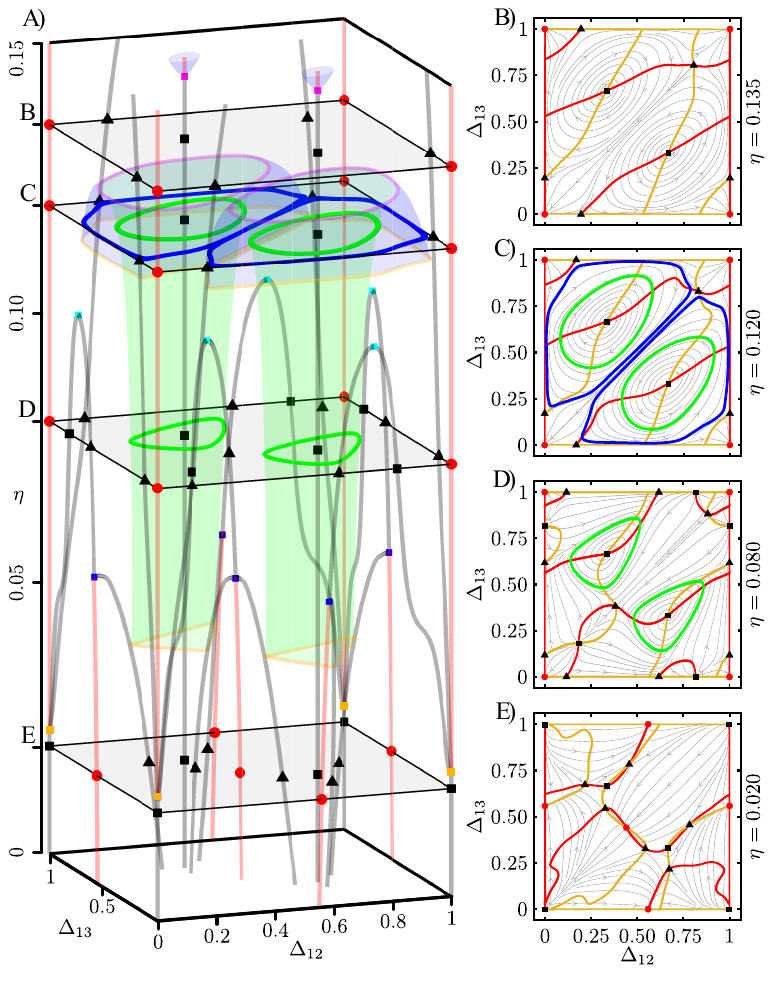}
    \caption{Bifurcation analysis of the phase-difference model associated with the network of FHN neurons.
    (A) Bifurcation diagram for $\eta \in [0,0.15]$. 
    Stable (unstable) limit cycles (LCs) form green (blue) surfaces, corresponding to branches of periodic orbits obtained by continuation in $\eta$. Black/red curves represent unstable/stable fixed points, cyan and blue squares represent saddle-node and pitchfork bifurcations, respectively, while the orange squares are associated with a complex series of local bifurcations occurring in a narrow parameter range, specifically comprising three saddle-nodes followed by two transcritical bifurcations occurring at the same parameter value. 
    The planes labeled B, C, D, and E section the bifurcation diagram for $\eta = 0.135,\,0.120, \,0.080$ and $\,0.020$, respectively. These planes correspond to the phase planes presented in panels B-E.
    In these planes, black squares, black triangles, and red circles correspond to unstable, saddle-type, and stable FPs, respectively, while the green/blue closed curves represent stable/unstable LC. (B--E) Phase planes corresponding to the sections presented in panel A. The grey lines indicate the local flow of the phase-difference model, while the red/orange lines are the $\Delta_{12}$/$\Delta_{13}$ nullcline. The remaining symbols and curves are as those presented in panel A.}
    \label{Fig1_FHN3}
\end{figure}

Figure~\ref{Fig1_FHN3}A 
shows the bifurcation diagram for \eqref{eq:dDelta} in the case of the FHN model, equations \eqref{dV_FHN} and \eqref{G_FHN}, for relatively small delays, $0\leq \eta\leq 0.15$,
in $(\Delta_{12},\Delta_{13},\eta)$ space.
Green and blue surfaces represent stable and unstable limit cycles (LCs), respectively, gray curves indicate unstable equilibria whereas the red curves are branches of stable equilibria.
The orange solid squares represent an intricate series of local bifurcations concerning three saddle-node 
bifurcations followed by two transcritical bifurcations occurring practically at the same parameter value, the solid blue squares represent simple pitchfork bifurcations (PF), and finally the cyan squares are saddle-node bifurcations. 

The four horizontal planes, indicated on the $\eta$ axis by B--E, correspond to the phase planes shown in the corresponding panels B--E. 
In these planes,  red dots and black solid squares represent stable and unstable fixed points (FPs), respectively, whereas solid black triangles indicate saddle points. 
The thin gray trajectories with arrows in panels 
B--E show representative flow directions in the phase-difference plane. The red and yellow curves indicate the nullclines of the phase-difference system. 

As the time delay increases from bottom to top, the system's dynamic behavior changes significantly.
For short delays, e.g. $\eta=0.020$ (Figure~\ref{Fig1_FHN3}E), the 
system is dominated by three stable equilibria 
at approximately $(0,1/2)$, $(1/2,0)$, and $(1/2,1/2)$.
The traveling wave (TW) FPs at $(1/3,2/3)$ and $(2/3,1/3)$, also known as splay states where subsequent neurons are shifted by a constant phase difference, are characterized by complex-conjugate eigenvalues with positive real parts, and are thus repelling. Trajectories spiral outward in a clockwise direction 
and are ultimately attracted to one of the stable equilibria. 

As the delay increases to $\eta = 0.080$ (Figure~\ref{Fig1_FHN3}D), the 
stable equilibria at approximately $(0,1/2)$, $(1/2,0)$, and $(1/2,1/2)$
lose stability in PFs 
at $\eta\approx 0.05$.
In addition, two stable (green) LCs emerge in heteroclinic bifurcations (orange LCs) at $\eta\approx 0.04$ and become the dominant rhythm, such that outward-spiraling trajectories are ultimately attracted to these stable patterns, which, however, coexist with a stable synchronous FP at $(0,0)$. 
When the delay further increases to $\eta = 0.120$ (Figure~\ref{Fig1_FHN3}C),  the stable LCs are still the dominant attractors, but they are now enclosed by unstable LCs born in a heteroclinic bifurcation at $\eta\approx 0.115$.
Finally, at $\eta = 0.135$ (Figure~\ref{Fig1_FHN3}B), the LCs have disappeared in saddle-node-of-periodic bifurcations (SNPOs), leaving the synchronous FP as the sole attractor of the system.

Overall, the bifurcation diagram calculated for the phase model corresponds to the previously proposed scenario \cite{qie2025synaptic}, but provides more details and reveals some differences, in particular regarding the birth of the stable limit cycle in the phase model, corresponding to slow modulation of the FHN oscillators. 
Further, since the equilibria are found as intersections of nullclines and followed by numerical continuation we obtain a complete set of FPs. In addition,  their stability is calculated directly, rather than being inferred from numerically calculated flows, guaranteeing a richer and more accurate description of the dynamics of the phase difference model.
\begin{figure}[tbp]
    \centering
    \includegraphics
    {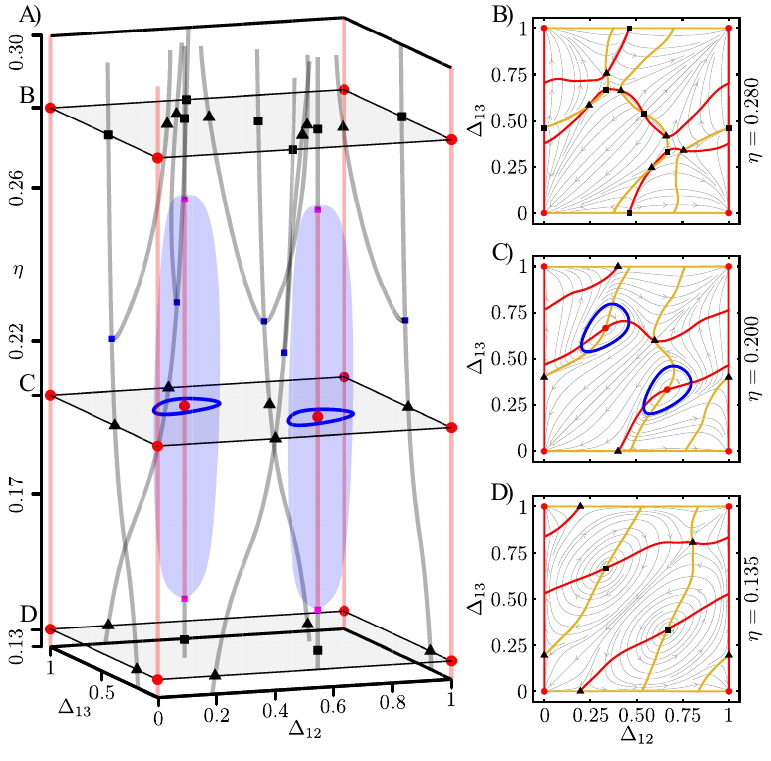}
    \caption{Bifurcation analysis of the phase-difference model associated with the network of FHN neurons (continued). (A) The bifurcation structure for $\eta \in [0.135,0.30]$.
    All symbols and curves are coded as in Figure~\ref{Fig1_FHN3}. Panels (B–D) present the phase portrait at selected delay values $\eta = 0.280$, $0.200$ and $0.135$, respectively, corresponding to the sections of the bifurcation diagram in panel A.}
    \label{Fig2_FHN3}
\end{figure}

Figure \ref{Fig2_FHN3} showcases an extension of Figure~\ref{Fig1_FHN3} for longer time-delays $\eta \in [0.135,0.30]$. 
We can see that the TWs first regain, and then lose, their stability in Hopf bifurcations (HBs) as the $\eta$-parameter is increased. 
Whenever the TWs are stable, as in Figure~\ref{Fig2_FHN3}C, unstable LCs consistently surround them, and only trajectories inside these cycles converge to the TWs. 
For all considered delays in Figure~\ref{Fig2_FHN3}, the synchronous FP remains stable with a large basin of attraction. 
When the TWs become unstable again at larger delays, trajectories reach the synchronous FP more directly than at shorter delays due to the presence of saddles,
compare Figure~\ref{Fig2_FHN3}D and Figure~\ref{Fig2_FHN3}B.

\begin{figure}[tbp]
    \centering
    \includegraphics
    {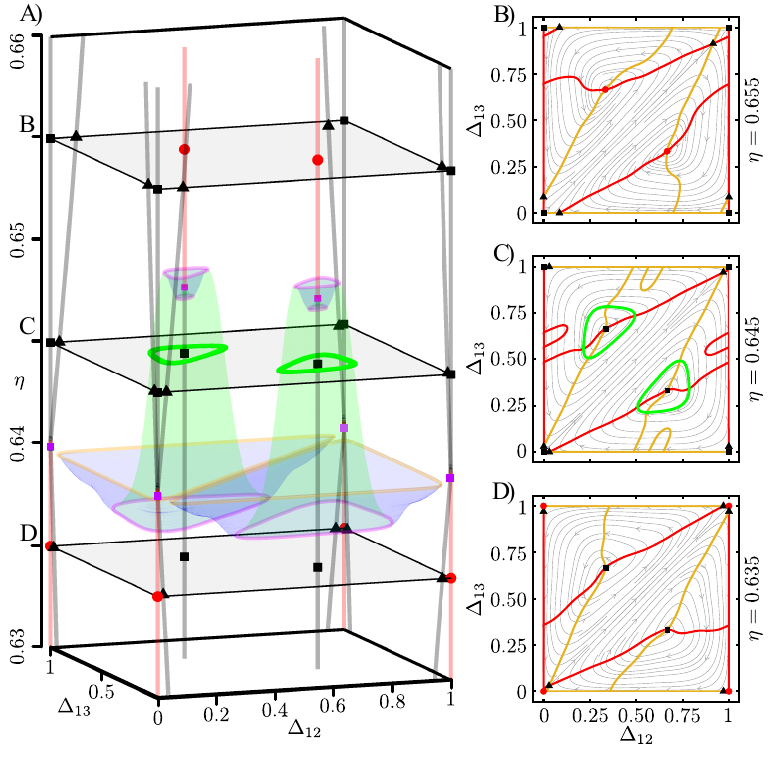}
    \caption{Bifurcation diagram and phase-plane analysis for the FHN phase-difference model for larger delays. The notation follows 
    Figure~\ref{Fig1_FHN3}. (A) Bifurcation diagram for $\eta \in [0.63,0.66]$. The purple square represents two transcritical bifurcations occurring for the same $\eta$ value. Panels (B–D) show the phase portraits corresponding to the three planes of panel A located at $\eta = 0.655$, $0.645$,
    and $0.635$, respectively.}
    \label{Fig3_FHN3}
\end{figure}
Figure~\ref{Fig3_FHN3} shows the bifurcation structure for $\eta$ between 0.63 and 0.66 (no limit cycles or interesting bifurcations occur for $0.30<\eta<0.63$, i.e., between Figures~\ref{Fig2_FHN3} and \ref{Fig3_FHN3}). 
When the two TW foci (black squares) are unstable in Figure~\ref{Fig3_FHN3}D, the synchronous FP at the origin is stable and determines the only robust rhythm.
As the delay parameter $\eta$ increases, the SNPOs lead to the birth of stable LCs around the unstable TWs and two larger unstable LCs that disappear in heteroclinic bifurcations at $\eta\approx 0.64$. 
At $\eta=0.645$ (Figure~\ref{Fig3_FHN3}C), the two stable LCs (green) solely determine the repertoire of the system, and all trajectories converge to these rhythmic patterns. The stable LCs eventually disappear in SNPOs when they encounter the unstable LCs born in the HBs, where the TWs regain stability. 

The synchronous FP loses stability through a double transcritical bifurcation occurring at $\eta \approx 0.64$ (purple squares), which is reflected in the differences seen near the origin between panels C and D of Figure~\ref{Fig3_FHN3}. 
For values of $\eta$ before the bifurcation, such as Figure~\ref{Fig3_FHN3}D, the synchronous FP is stable and surrounded by three saddles (because the map is periodic, two of the saddles are located close to the top right corner). 
For values of $\eta$ past the bifurcation, e.g., Figure~\ref{Fig3_FHN3}C, the FPs close to the origin survive with the origin losing its stability while the saddles remain saddles by switching their attracting and repelling eigendirections.

\subsection{The slow component in higher-dimensional FHN networks}
An advantage of our approach based on phase-differences and numerical construction of bifurcation diagrams is that it is immediately applicable to larger networks.
Therefore, we extend the phase-difference model to higher-dimensional systems consisting of $N$ neurons to investigate if slow delay-induced rhythms also appear in larger networks. Specifically, we consider ring networks composed of $N=5,7,$ or $9$ neurons.

\begin{figure}[htbp]
    \centering
    \includegraphics
    {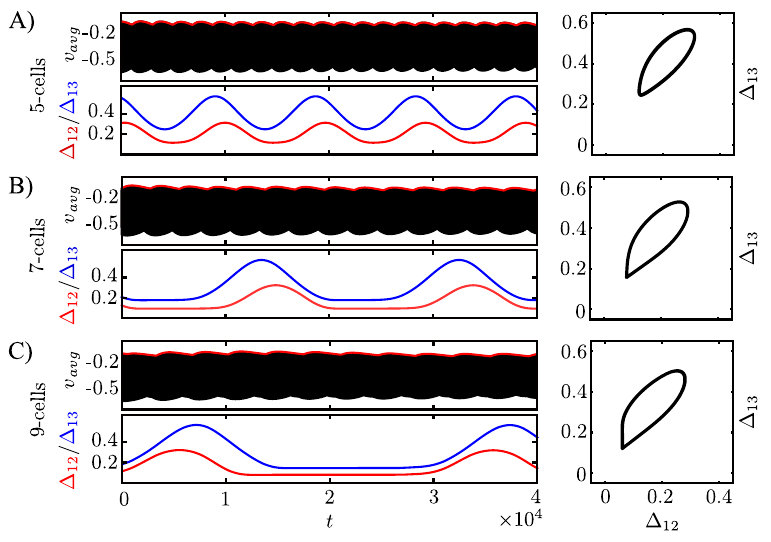}
    \caption{Slow modulation in larger networks of FHN oscillators. Each panel illustrates the temporal evolution of the average voltage $v_{avg}$, its envelope (red curve), the phase differences between neuron 1 and neuron 2 or 3 ($\Delta_{12}$/$\Delta_{13}$), and the limit cycle projected onto the $(\Delta_{12},\Delta_{13})$ phase plane. Panels (A)/(B)/(C) show simulations for a network with 
    of 5/7/9 neurons, respectively, and with $\eta=0.15$.}
    \label{fig:fhnN}
\end{figure}

Since the thorough bifurcation analysis of three-cell network showed that LCs arose only in the vicinity of the TWs ($\Delta_{1j} = \frac{j-1}{N}, j=2,\dots,N$), we decided to focus on the 
the analysis of the dynamics around these FPs.
Figure~\ref{fig:fhnN} shows that stable LCs can emerge in these higher-dimensional systems. Moreover, the slow envelope can be identified in the average membrane potential (Figure~\ref{fig:fhnN}ABC, upper panels), indicating the presence of slow modulation of the fast oscillatory dynamics.
As the number of neurons increases, with the chosen value of the delay, the LC approaches a saddle-point as seen from the increasingly pointed orbit (Figure~\ref{fig:fhnN}ABC, right panels), and the increasingly long time that the system stalls at low $\Delta_{12}$ and $\Delta_{13}$ values (Figure~\ref{fig:fhnN}ABC, lower panels).
Consequently, more time is required to complete one revolution, leading to an increase in the period of the LCs around TWs. 

\section{Extension to the Morris–Lecar model with delay synaptic coupling}
\label{sec:ML}
Although the FitzHugh–Nagumo system provides a simplified prototype of excitable dynamics, it remains important to verify whether the observed rhythmic organization persists in more biophysical models.
In this section, we apply computational simulation, phase space analyses and bifurcation theory to (i) unveil the presence of the slow rhythm, (ii) understand its origin and (iii) how it changes as the delay $\eta$ is varied.
Here, we investigate a circuit composed of three identical, Morris–Lecar (ML) oscillators with delayed, inhibitory, and weak synaptic coupling, following the formulation of Rinzel and Ermentrout \cite{rinzel1998analysis}. For each neuron $(j \in \{1,2,3\})$, the membrane potential $V_j$ and the gating variable $w_j$ evolve according to
\begin{equation}
    \begin{aligned}
    C_m \frac{d{V}_j}{dt} &= -I_L(V_j)-I_K(V_j,w_j)-I_{Ca}(V_j)+I\,+\,g\sum_{\substack{i=1 \\ i \neq j}}^{3}\,G_{ML}(V_j(t),V_i(t-\rho)), \\
    \frac{d{w}_j}{dt} &= \bigl[w_\infty(V_j) - w_j\bigr]\, \tau_w^{-1}(V_j),\\
    \end{aligned}
    \label{eq:ML_3cell_delay}
\end{equation}
where $I_x$ with $x \in \{L, K, Ca\}$ is an ionic current, and  $C_m = 20\ \mu$F/cm\textsuperscript{2} is the membrane capacitance. The term $g$ denotes the synaptic coupling strength, while $\rho$ is the synaptic transmission delay, which we recall to be defined as $\rho = \eta\, T$, with $T$ the period of the single oscillator and $\eta$ the fractional delay. The currents are, respectively, a leakage-type ($I_L$), potassium ($I_K$), and calcium ($I_{Ca}$) currents. They follow the equations,
\begin{equation}
    \begin{aligned}
        I_L(V_j) &= g_{\mathrm{L}}(V_j-E_{\mathrm{L}}), \\
        I_K(V_i,w_j) &= g_{\mathrm{K}}\, w_j\,(V_j - E_K), \\
        I_{Ca}(V_j) &= g_{\mathrm{Ca}}\, m_\infty(V_j)\,(V_j - E_{\mathrm{Ca}}). \\ 
    \end{aligned}
\end{equation}
The $g_x$ terms, with $x \in \{L, K, Ca\}$, represent the maximal conductances, while $E_x$ is the corresponding reversal potential. 
The activation of the Ca\textsuperscript{2+} currents is assumed to be instantaneous, while the gating variable of the K\textsuperscript{+} channel ($w_j$) is governed through a first-order ODE. Its steady-state ($w_{\infty}$) and time scale ($\tau_{w}$) 
functions read,
\begin{equation}
   \begin{aligned}
       m_\infty(V) &= \frac{1}{2}\left[1 + \tanh\!\left(\frac{V - V_{m}}{s_m}\right)\right], \\
        w_\infty(V) &= \frac{1}{2}\left[1 + \tanh\!\left(\frac{V - V_w}{s_w}\right)\right], \\
        \tau_w(V) &= \Bigg[\cosh\!\left(\frac{V - V_w}{2s_w}\right)\Bigg]^{-1}.
    \end{aligned}
\end{equation}
Finally, we assume the coupling function to be
\begin{equation}
        G_{ML}(V_j,V_i) = s\!\left(V_i\right)\:(E_{\mathrm{SYN}}-V_j),
\end{equation} 
where $E_{\mathrm{SYN}}$ is the reversal potential of the synapse and it is modeled according to a typical GABAergic synapse in the CNS, as suggested in~\cite{ermentrout2010mathematical}. The function $s(V)$, which models the synaptic activation, is a smooth sigmoidal function of the form,
\begin{equation}
    s(V_i) = \frac{1}{2}\left[1 + \tanh(\,V_i/12)\right].
\end{equation}
The maximal conductances, $g_L$, $g_{Ca}$, $g_K$, and $g$, are set to 2, 4.4, 8, and 0.0005 mS/cm\textsuperscript{2}, respectively. The Nernst potentials
$E_L$, $E_{Ca}$, $E_{K}$, and $E_{SYN}$ are -60, 120, -84, and -70 mV, respectively. The parameters of gating variable, steady-state and time scale functions, 
$V_m$, $s_m$, $V_w$, and $s_w$,
are -1.2, 18, 2, and 30 mV. The current $I$ applied to each neuron of the network is equal to 100 $\mu$A/cm\textsuperscript{2}.
With these parameters, an isolated neuron ($g=0$ mS/cm\textsuperscript{2}) exhibits a stable periodic oscillation of period $T \approx 85.24\  \mathrm{ms}$. 
This rhythmic activity provides the dynamical foundation for the phase reduction and delay-induced rhythm analysis.

\begin{figure}[htbp]
    \centering
    \includegraphics
    {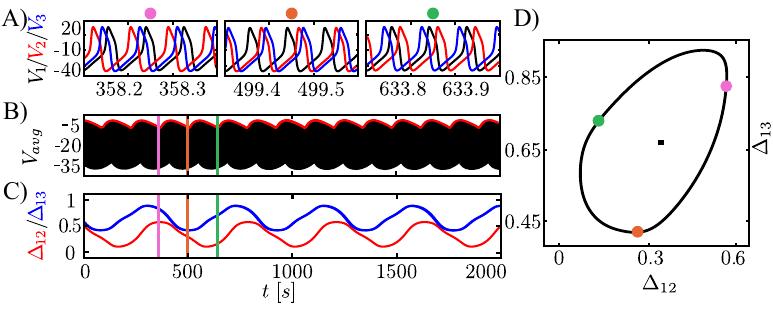} 
    \caption{Slow modulation in a network of Morris-Lecar neurons with delayed, inhibitory coupling.
    Panels (A), (B), and (C) show the time series of $V_j$ with j $\in \{1,2,3\}$, $V_{\mathit{avg}}$, and $\Delta_{12}, \Delta_{13}$ for $\eta=0.21$. (D)
    The corresponding evolution of the phase-difference model in the $(\Delta_{12},\Delta_{13})$ phase plane.
    Notation as in Figure~\ref{Fig1_FHN3_TS}.
    }
    \label{Fig3_ML3_TS}
\end{figure}

Numerical simulations, using the ML neuronal network and the corresponding phase-difference models, reveal a slow rhythm due to modulation of the single-unit oscillations,
similarly to what presented for the FHN model. Specifically, for $\eta=0.21$, as shown in Figure~\ref{Fig3_ML3_TS}AB, the presented time series demonstrates that this slow component carries over to 
more biophysical networks of neurons. 
The period of the slow modulation is three orders of magnitude slower than $T$.
As for the FHN model, the oscillatory behavior corresponds to a cyclic evolution in the phase-difference variables $\Delta_{12}$ and $\Delta_{13}$ (Figure~\ref{Fig3_ML3_TS}CD).

\begin{figure}[!h]
    \centering
    \includegraphics
    {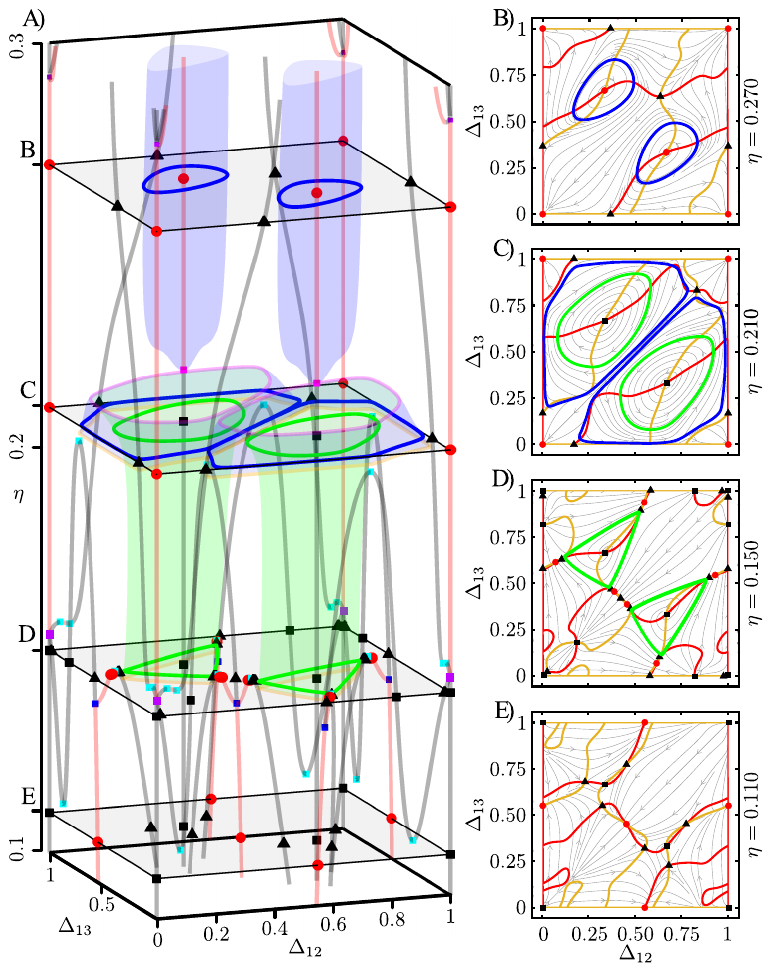}
    \caption{Bifurcation analysis of the phase-difference model corresponding to the Morris-Lecar model.
    Notation as in Figure~\ref{Fig1_FHN3}. (A) Bifurcation diagram for $\eta \in [0.10,0.30]$. Panels (B–D) show the phase portrait for $\eta = 0.110$, $0.150$, $0.210$ and $0.270$, respectively.}
    \label{Fig1_ML3}
\end{figure}

We now aim to explore the bifurcation structure leading to the generation of these slow components.
Figure~\ref{Fig1_ML3} shows the bifurcation diagram as the delay is changed in the ML phase-difference model \eqref{eq:ML_3cell_delay} for relatively short delays ($\eta\in \left[0.1, 0.3\right]$).
For small delay values (e.g., $\eta=0.11$, Figure~\ref{Fig1_ML3}E), three stable equilibria, located at $(0,1/2)$, $(1/2,0)$, $(1/2,1/2)$ approximately (which are similar to those detected in Figure~\ref{Fig1_FHN3}A), dominate the dynamics, with their basins of attraction 
completely determined by the six saddles surrounding the two unstable TWs.
For a value of $\eta$ close to 0.142, the stable FPs lose their stability via PFs, from where two branches of stable FPs appear. These branches move towards the central saddles as $\eta$ increases. Before the saddles disappear in saddle-node bifurcations (cyan dot in Figure~\ref{Fig1_ML3}A), two heteroclinic bifurcations occur (orange LCs in Figure~\ref{Fig1_ML3}A, close to the D section). These bifurcations give rise to two stable LCs.
For a small $\eta$ interval (approximately, $\eta \in [0.148, 0.151]$), both the stable LCs and the stable FPs coexist, yielding bistability. 
Figure~\ref{Fig1_ML3}D illustrates this condition; the trajectory converges towards the stable LC or a stable FP, depending on the initial conditions. The boundaries of the basin of attraction are again organized in this case, through the nearby saddles.

As $\eta$ increases, the synchronous FP stabilizes through
a double transcritical bifurcation occurring at $\eta$ equal to 0.153 (purple square in Figure~\ref{Fig1_ML3}A). For values of $\eta$ higher than 0.153 the phase-difference model presents bistability between the 
stable synchronous FP and the stable LCs. Once again, the three saddles surrounding the TWs partition the phase space into different basins of attraction. For the value of $\eta$ close to 0.208, two unstable LCs arise from heteroclinic bifurcations (orange LCs, Figure~\ref{Fig1_ML3}A). As observed in the case of the FitzHugh-Nagumo network of neurons, the unstable LCs act as a separatrix between the basin of attractions of the stable LCs and the stable synchronous FP. When $\eta$ is increased, the stable and the unstable LCs collide in two SNPOs bifurcations occurring at $\eta$ equal to 0.217 (magenta LCs, Figure~\ref{Fig1_ML3}A). Past this mechanism of LCs disappearance, for $\eta \in [0.217, 0.222]$, the only stable attractors are the synchronous FP. When $\eta$ is equal to approximately 0.222, the TWs gain stability via subcritical HBs (magenta squares in Figure~\ref{Fig1_ML3}A). From the HBs a family of unstable LCs arises and acts as a separatrix between the basin of attraction of the stable TWs and synchronous FP, as illustrated in Figure~\ref{Fig1_ML3}B. For values of $\eta$ close to 0.325, higher than those presented in Figure~\ref{Fig1_ML3}A, the unstable LCs disappear through a heteroclinic connection. 
When $\eta$ is increased further, the stable TWs lose stability via a second subcritical HBs.

\begin{figure}[tbp]
    \centering
    \includegraphics
    {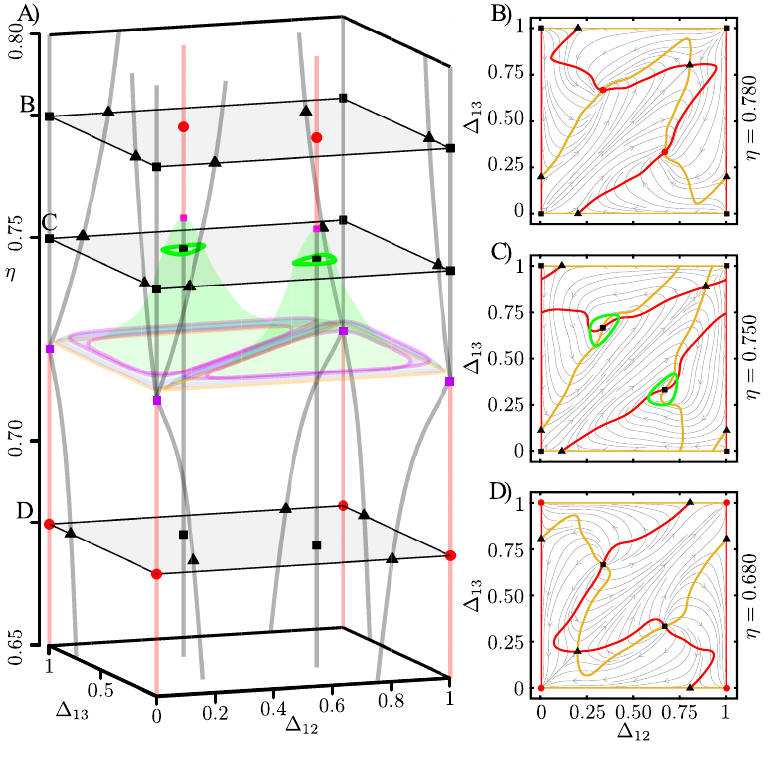}
    \caption{Bifurcation analysis of the phase-difference model corresponding to the Morris-Lecar model (continued). The organization of this figure follows
    Figure~\ref{Fig1_FHN3}. (A) Bifurcation diagram for $\eta \in [0.65,0.80]$. Panels (B-D) show the phase plane, nullclines, FPs, and flow of the phase-difference model at $\eta = 0.680$, $0.750$, and $0.780$, respectively.}
    \label{Fig2_ML3}
\end{figure}

Figure~\ref{Fig2_ML3} illustrates the bifurcation structure
for values of $\eta \in [0.65,0.80]$. The stable synchronous state, which is the only dominant rhythm in Figure~\ref{Fig2_ML3}D, loses stability at $\sim$ 0.72
via a double transcritical bifurcation (purple square in Figure~\ref{Fig2_ML3}A). As the delay continues to increase, 
stable LCs appear through a heteroclinic bifurcation. 
After a small interval of $\eta$ where the phase-difference model shows bistability between different stable LCs, the branch of unstable and stable LCs coalesce in SNPOs (magenta LCs in Figure~\ref{Fig2_ML3}A). For larger $\eta$, the TWs undergo a third HB, regain stability, becoming the only stable attractor in the phase plane.
 
\section{Delay-induced slow rhythms in a Next-Generation Neural Mass Model}
To further validate that the phase-difference framework developed is not restricted to relaxation oscillators (FHN, Section~\ref{sec:FHN}) or two-dimensional conductance-based models (ML, Section~\ref{sec:ML}), we next consider a three-dimensional next-generation neural mass model describing synaptically coupled, inhibitory quadratic integrate-and-fire (QIF) neurons \cite{ceni2020cross}.
Rather than simulating a high-dimensional spiking network, this model was derived as the exact mean-field reduction
introduced by Montbrió et al.~\cite{montbrio15} to provide a closed, low-dimensional description of macroscopic population dynamics.

We consider three interacting neuronal populations indexed by $j\in\{1,2,3\}$. For each population $i$, the macroscopic state is characterized by the firing rate $r_j(t)$, the mean membrane potential 
$v_j(t)$, and the mean synaptic activity $s_j(t)$. Synaptic transmission delays are incorporated 
by assuming that the input from population $j$
arrives at population $i$ after a fixed delay 
$\rho$ \cite{devalle2018dynamics,Pazo2016,chen2024population}.

The mean-field (MF) dynamics with delayed coupling is given by
\begin{equation}
\label{eq:MPR_3pop_delay}
    \begin{aligned}
    \frac{dr_j}{dt}  &= \frac{\delta}{\tau_m^2\pi} + \frac{2\,r_j\,v_j}{\tau_m}, \\
    \frac{dv_j}{dt}  &= \frac{v_j^2 + \bar{\xi}}{\tau_m} + J_s\, 
    s_j - \tau_m\big(\pi r_j\big)^2 + J_c \sum_{\substack{i=1 \\ i \neq j}}^{3}
    \, G_{MF}\left(s_i\!\left(t-\rho\right)\right), \\
    \frac{ds_j}{dt} &= \frac{1}{\tau_d}\big(-s_j+ r_j\big),
    \end{aligned}
\end{equation}
for $j=1,2,3$. In this latter case, the coupling function is a simple sign change, namely $G_{MF}(s_i)=-s_i$. Here $\tau_m=10$ ms is the membrane time constant and $\tau_d=8$ ms is the synaptic time constant.
The parameters $\bar\xi=0.05$ and $\delta=0.05$ denote the center and half-width of the Lorentzian distribution of intrinsic neuronal excitabilities in population $j$
, respectively.
Intra-population inhibitory coupling is modeled by the self-coupling coefficient $J_s=-20$, while the cross-coupling coefficients $J_c =0.0005$ describe the strength of the inhibitory synaptic coupling of population $i$ acting on population $j$.
The delay $\rho$ models synaptic transmission latencies. 

\begin{figure}[htbp]
    \centering
    \includegraphics
    {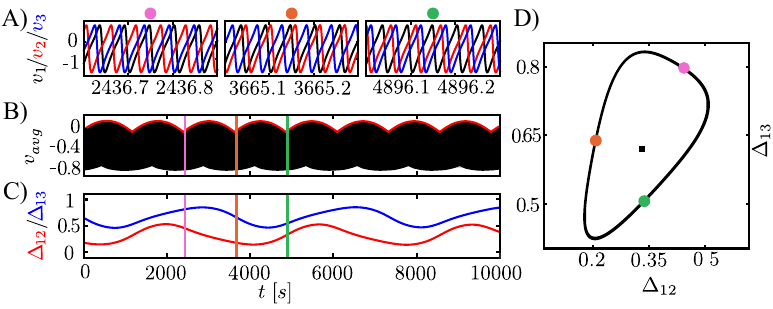} 
    \caption{Slow modulation of the next-generation mean-field model. Notation as in Figure~\ref{Fig1_FHN3_TS}. The time series of $v$, $v_{\mathrm{avg}}$, and $\Delta_{12}, \Delta_{13}$ in the MF model with $\eta=0.43$ are shown in panels (A), (B), and (C), respectively. (D)  The corresponding phase plane of $(\Delta_{12},\Delta_{13})$.}
    \label{Fig3_MF3}
\end{figure}

Simulations of the MF network model for $\rho\approx 23.6$ ms ($\eta=0.43$) show that the slow modulation of the single-unit activity is also present in this population model (Figure~\ref{Fig3_MF3}AB), illustrating the generality of the observed slow modulation phenomenon.
The delay-induced slow rhythm is also captured by the phase-difference model, as seen in its temporal dynamics and phase-plane representation (Figure~\ref{Fig3_MF3}CD).

As in the earlier sections of the paper, we investigate the bifurcation structure leading to the generation of the observed slow oscillations using the phase-difference model.
Rather than illustrating the full bifurcation structure of the system, we present only the $\eta$ intervals where the phase-difference model associated with the MF network presents LCs.

\begin{figure}[htbp]
    \centering
    \includegraphics
    {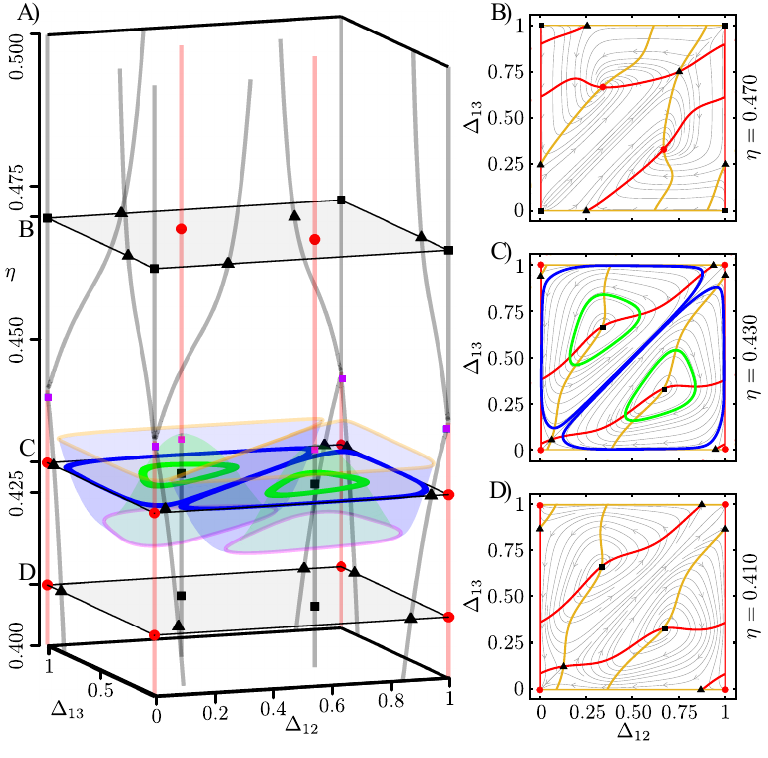}
    \caption{Bifurcation analysis of the phase-difference model corresponding to the mean-field model.
    Notation follows as in Figure~\ref{Fig1_FHN3}. (A) Bifurcation diagram of a network of three mean-field models for $\eta \in [0.4, 0.5]$. (B-D) Phase plane analyses results associated with the sections labelled B, C, and D illustrated in panel A.}
    \label{Fig1_MF3}
\end{figure}

Figure~\ref{Fig1_MF3} shows the bifurcation diagram 
for values of the fractional delay between 0.4 and 0.5. For $\eta$ smaller than 0.42, the only stable attractor of the system is the synchronous FPs (Figure~\ref{Fig1_MF3}D) located at the four corners. 
For $\eta \in [0.422,0.435]$ approximately, the phase-difference model shows a pair of stable and unstable LCs arising from two SNPOs (magenta LCs in Figure~\ref{Fig1_MF3}A). In this parameter interval, the unstable LCs separate the basin of attraction of the stable synchronous FP and the stable LCs. In fact, as
seen in Figure~\ref{Fig1_MF3}C, only those gray trajectories inside the unstable LCs (blue closed curve) converge towards the inner stable LCs (green closed curve).

\begin{figure}[htbp]
    \centering
    \includegraphics
    {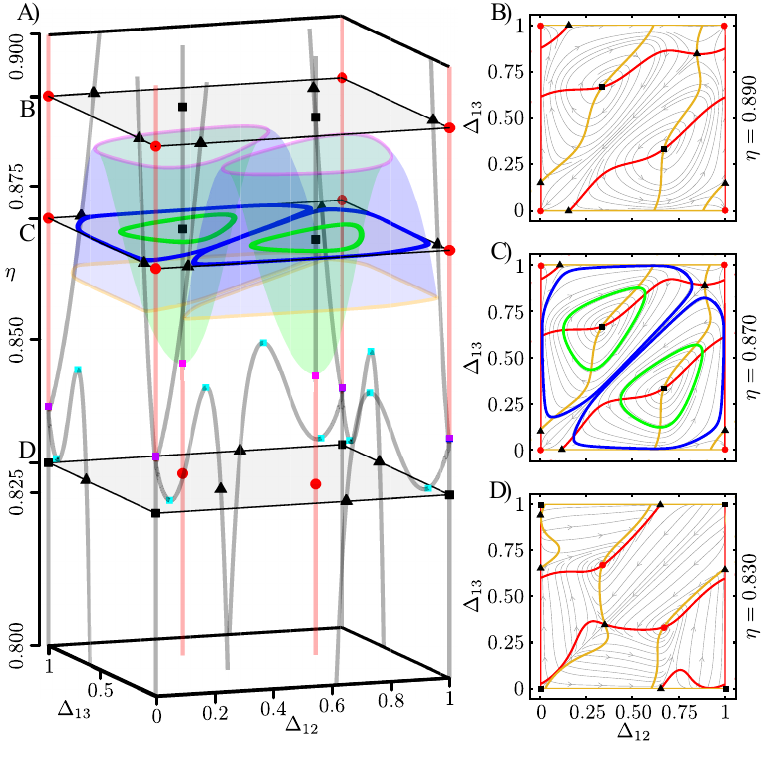}
    \caption{Bifurcation analysis of the phase-difference model corresponding to the mean-field model (continued). The organization of the figure follows those presented previously. (A) Bifurcation diagram for values of $\eta \in [0.8, 0.9]$. (B-D) Phase planes corresponding to the sections of the bifurcation diagram in panel A. 
    }
    \label{Fig2_MF3}
\end{figure}

When $\eta$ increases, the unstable LCs disappear in a heteroclinic connection (orange LCs in Figure~\ref{Fig1_MF3}A), while the stable LCs disappear when the TWs gain stability in supercritical HBs
(magenta squares). 
Following a double transcritical bifurcation (purple square in Figure~\ref{Fig1_MF3}A), which makes the synchronous solution unstable, the TWs are the only stable attractors of the system up to $\eta=0.835$, as presented in Figures~\ref{Fig1_MF3}B, \ref{Fig2_MF3}D, and~\ref{Fig2_MF3}A. 
After this value of $\eta$, the synchronous solution regains stability through a double transcritical bifurcation (purple square in Figure~\ref{Fig2_MF3}A). 
For values of $\eta$ close to 0.85, the stable TWs lose their stability via supercritical HBs, from which stable LCs arise (Figure~\ref{Fig2_MF3}C). For larger delays, the branch of stable LCs disappear in SNPOs when encountering a branch of unstable LCs (blue closed curve in Figure~\ref{Fig2_MF3}C) arising from heteroclinic bifurcations (orange LCs, Figure~\ref{Fig2_MF3}A). 
After this bifurcation, the synchronous solution is the only stable attractor of the system (Figure~\ref{Fig2_MF3}B).

\section{Discussion}
Slow brain rhythms are important for cognitive processes and memory consolidation.
In this work, we investigated the effects of synaptic transmission delays on phase dynamics in ring networks of coupled neurons, showing that slow modulatory rhythms generically emerge. Based on these results, it is tempting to speculate that delays in neural transmission are at least partly involved in the emergence of slow brain rhythms.

To dissect the underlying mechanisms of the observed slow rhythms induced by delayed coupling, we derived a phase-difference system, based on phase response curves, where the delay appears as a standard parameter.
The resulting phase system is therefore naturally suited for bifurcation analysis, allowing a systematic characterization of the existence, stability, and transitions of slow periodic solutions.

In the dimensionless FitzHugh–Nagumo model, we identified
a clear slow component, which manifests as a low-frequency envelope in the time series of the average voltage, and corresponds to continuous phase shifts in the voltage activity of individual neurons. Furthermore, we performed a systematic bifurcation analysis by treating the synaptic delay as a control parameter using the phase model. 
This reveals the dynamical mechanisms by which limit cycles emerge near traveling-wave solutions (FPs in the phase-difference model).
We extended the analysis to higher-dimensional ring networks and identified stable limit cycles, indicating that the phenomenon of a delayed-coupling-induced slow component persists beyond low-dimensional networks.

To verify that this phenomenon is not restricted to simplified models, we then considered the more biologically realistic Morris–Lecar model. 
Despite its more detailed conductance-based structure, we observed similar stable limit cycles in the phase-difference model, corresponding to slow modulation of the single-unit dynamics in the original model. Further,
we extended the analysis to a mean-field neuronal model based on population dynamics, thereby moving from single-neuron descriptions to collective behavior. 
Within this higher-level framework, stable limit cycles and
slow modulation are likewise observed. These results demonstrate that the emergence of the slow modulation driven by phase dynamics as a result of delayed coupling persists across different levels of neuronal modeling, highlighting its robustness and generality.

\begin{figure}[htbp]
    \centering
    \includegraphics
    {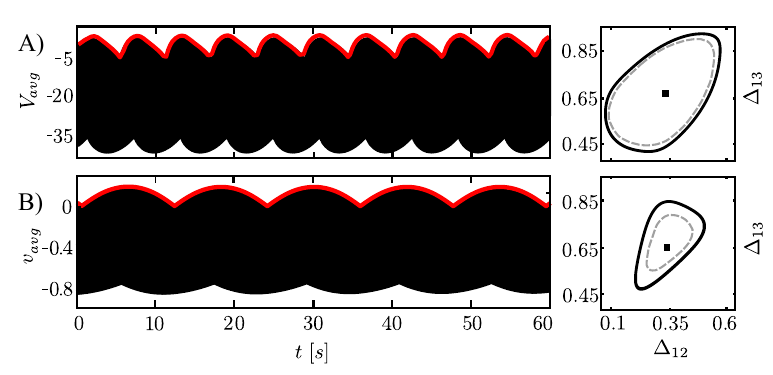} 
    \caption{The period of the slow modulation is lower with larger coupling strengths.
    (A) Left: Time series of the average voltage (black) for the ML network with $g=0.01$, $\eta=0.21$  and the slow envelope of the average voltage (red). Right: Phase-difference phase plane showing the limit cycle of the full ML model (gray dashed curve) and of the corresponding phase model (black solid curve).
    (B) As in panel A for the MF network with $J_c=0.05$, $\eta=0.43$.}
    \label{Fig_larger_g}
\end{figure}

The use of more realistic models allows the comparison of the period of the simulated slow modulatory patterns with those of experimentally observed brain rhythms. 
The delta band corresponds to the slowest neural oscillations typically identified in EEG recordings, and has a frequency of $0.5-4$ Hz, i.e., a period of at most a few seconds. In contrast, our simulations in 
Figures~\ref{Fig3_ML3_TS} and \ref{Fig3_MF3} show slow oscillations with a period of several minutes. 
The origin of these extremely low frequencies is the low value of $g$, the coupling strength, as can be seen as follows. On the one hand, weak coupling guarantees that the phase-difference model is a good approximation to the full model. On the other hand, in the first-order approximation \eqref{eq:dDelta}, $g$ simply rescales time such that a small $g$ produces very slow phase-difference dynamics. Consequently, to obtain slow patterns with periods of a few seconds, we increase $g$ greatly, which produces much more realistic slow dynamics at the expense of a larger discrepancy between the full model and the phase-difference model 
(Figure~\ref{Fig_larger_g}). 
Nonetheless, we believe that the insight gained from the bifurcation analyses of the phase-difference models, valid for low $g$ values, provides insight into the dynamical origin also of the slow rhythms with more realistic periods obtained for larger values of $g$. 
Including higher-order terms in \eqref{eq:dDelta} could improve the accuracy of the phase-difference model for larger coupling strengths.

\begin{figure}[htbp]
    \centering
    \includegraphics
    {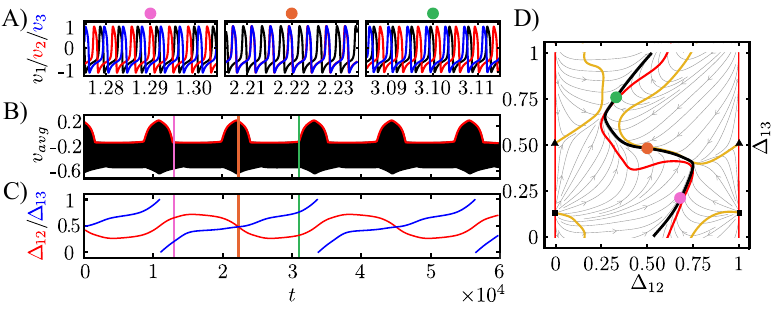} 
    \caption{Slow modulation in an asymmetric FHN network with delays only in the bidirectional coupling
    between neurons 1 and 2 $(\eta_{12}=0.8)$.
    Notation as in Figure~\ref{Fig1_FHN3_TS}. 
    The time series of $v$, $v_{\mathit{avg}}$, and $\Delta_{12}$, $\Delta_{13}$ are shown in panels (A), (B), and (C), respectively.
    (D) The corresponding phase plane of $(\Delta_{12},\Delta_{13})$. The black curve corresponds to a limit cycle on the torus. }
    \label{Fig2_FHN3_TS}
\end{figure}

A limitation of our study is that we have only treated symmetric systems with identical neural oscillators coupled symmetrically. Future studies could use the approach presented here to investigate systems where this symmetry is broken to systematically search for slow rhythms and describe the underlying dynamical mechanisms.
Interestingly, and a first step in this direction, we find that the slow modulation can also be observed in a bidirectional, asymmetric delayed FHN motif (Figure~\ref{Fig2_FHN3_TS}). 
In this case, only two synaptic connections are delayed, which breaks the symmetry of the system. 
Yet, the averaged voltage $v_{\mathrm{avg}}$ exhibits a clear slow modulation (Figure~\ref{Fig2_FHN3_TS}B).
The asymmetry in the corresponding phase-difference model leads to dynamics exhibiting a continuous phase shift, which corresponds to a limit cycle that can only be seen by treating the phase-space as a torus. Indeed, as shown in Figure~\ref{Fig2_FHN3_TS}D, the trajectory in the $(\Delta_{12},\Delta_{13})$ phase plane does not form a closed orbit. 
Instead, $\Delta_{13}$ continuously drifts and wraps around the periodic phase domain. 
Trajectories starting from different initial conditions are all attracted toward this invariant curve.
Similar results are found with asymmetry in the coupling strengths.

In conclusion, we have studied the mechanisms by which delayed, inhibitory coupling between neural oscillators can produce slow modulation that is not present within the single units and is much slower than the delays. Our findings and future extensions of our ideas to more realistic -- larger and heterogeneous -- networks could be relevant for understanding how slow delta rhythms may appear from delayed neural communication.

\appendix
\section{The purpose of phase reduction} 
Rhythmic activity is a common feature of neural systems, especially in networks where interacting oscillators generate coordinated phases.
In many coupled networks, the key question is not the precise waveform of each oscillator, but the relative phase among oscillators.
Detailed mathematical models capture oscillator dynamics and cell-to-cell interactions, allowing us to simulate how parameters, such as coupling strength and delays, together with network structure, shape the rhythms.
However, such models are often complex, involving multiple state variables. 
This motivates the use of phase reduction. It provides a low-dimensional representation of the dynamics and a tractable setting for bifurcation analysis. Focusing on relative phases clarifies how rhythmic states are organized, while remaining computationally efficient for large-scale neural networks.

\section{Bifurcation analysis and bifurcation diagrams}
Many biological systems change their behavior in response to variations in parameters. For example, quiescent neurons clamped with an external current can start firing when the stimulus becomes sufficiently strong. Identifying such critical parameter values typically requires multiple experiments and is crucial for determining the parameter ranges over which the qualitative behavior of the system remains unchanged. However, because this procedure can be labor-intensive, the analysis of computational models provides a valuable alternative.
Bifurcation analysis is used to identify so-called bifurcation points 
where the system changes behavior qualitatively, such as a neuron going from quiescence to action potential firing beyond a certain stimulus threshold.
Mathematically, bifurcations are mechanisms that explain the emergence, stability changes, or disappearance of fixed points and limit cycles in response to variations in one or more model parameters. The results of a bifurcation analysis are typically summarized in bifurcation diagrams, which indicate where bifurcations occur.
In this work, we use bifurcation analysis to demonstrate the existence of limit cycles in the phase-difference model and, consequently, to determine the range of the fractional delay parameter $\eta$ for which slow rhythms can be observed in 
inhibitory neuronal systems.

\begin{reading}
For readers interested in these fields, we recommend the following resources:
\begin{itemize}
    \item \textit{Phase reductions}: 
    The textbook \cite{ermentrout2010mathematical} contains a simple yet comprehensive introduction to phase reduction and phase response curves. For more advanced and specific materials on three-cell networks and delays in weakly coupled oscillators we suggest 
    \cite{qie2025synaptic,campbell2012phase,campbell2018phase}.
    \item \textit{Bifurcation theory}:
    Many excellent books providing a gentle introduction to 
    bifurcation analysis exist. We have enjoyed, among others, \cite{ermentrout2010mathematical,strogatz2024nonlinear,izhikevich2007dynamical}. 
    A more advanced and mathematical textbook is \cite{Kuznetsov2023}. For 
    researchers looking for resources to perform numerical bifurcation analysis, we suggest using XPPAUT~\cite{ermentrout02xppaut}.
\end{itemize}

\end{reading}

\section*{Acknowledgments}
X.~Qie  thanks the Department of Information Engineering, University of Padova for hosting her during
the visit, where the present work was carried out.

\section*{Author Contributions}
All authors have read and approved the final manuscript.  
In addition, the following contributions occurred:  Conceptualization: Morten Gram Pedersen; 
Methodology: Morten Gram Pedersen; Formal analysis and investigation: Xinxin Qie, Matteo Martin; 
Writing - original draft preparation: Xinxin Qie, Matteo Martin, Morten Gram Pedersen; Writing - review and editing: Xinxin Qie, Matteo Martin, Shenquan Liu, Morten Gram Pedersen; 
Funding acquisition: Xinxin Qie, Shenquan Liu, Morten Gram Pedersen; Resources: Shenquan Liu, Morten Gram Pedersen; Supervision: Matteo Martin, Morten Gram Pedersen.

\section*{Conflict of interest} 
The authors declare there are no conflicts of interest.

\section*{Data \& Code Availability Statement}
The code can be found at the following link, \url{https://researchdata.cab.unipd.it/id/eprint/1848}.

\bibliographystyle{siamplain}
\bibliography{references}

\end{document}